\documentclass[letter]{emulateapj}

\usepackage{amsmath, amssymb, epsfig}

\slugcomment{Submitted to Astrophys. J.}

\shorttitle{Dead calm areas in the quiet Sun}
\shortauthors{Mart\'\i nez Gonz\'alez et al.}

\begin{document}

\title{Dead calm areas in the very quiet Sun}

\author{M. J. Mart\' inez Gonz\'alez, R. Manso Sainz, A. Asensio Ramos}
\affil{Instituto de Astrof\' isica de Canarias, V\' ia L\'actea s/n, E-38205 La Laguna, Tenerife, 
Spain}
\affil{Departamento de Astrof\'\i sica, Universidad de La Laguna, E-38205 La Laguna, Tenerife, Spain}
\author{E. Hijano}
\affil{McGill Physics Department, 3600 rue University, Montr\'eal, QC H3A 2T8, Canada}

\begin{abstract}
We analyze two regions of the quiet Sun ($35.6\times 35.6$~Mm$^2$) observed at high spatial resolution ($\lesssim$100~km) in polarized light by the IMaX spectropolarimeter onboard the Sunrise balloon. 
We identify 497 small-scale ($\sim$400~km) magnetic loops, appearing at an effective rate of 0.25~loop~h$^{-1}$~arcsec$^{-2}$; further, we argue that this number and rate are underestimated by $\sim$30\%.
However, we find that these small dipoles do not appear uniformly on the solar surface: their spatial distribution is rather filamentary and clumpy, 
creating {\em dead calm} areas, characterized by a very low magnetic signal and a lack of organized loop-like structures at the detection level of our instruments,
that cannot be explained as just statistical fluctuations of a Poisson spatial process.
We argue that this is an intrinsic characteristic of the mechanism that generates the magnetic fields in the very quiet Sun. The spatio-temporal coherences and the clumpy structure of the phenomenon suggest a recurrent, intermittent mechanism for the generation of magnetic fields in the quietest areas of the Sun.
\end{abstract}


\maketitle

\section{Introduction}

During the last few years, our understanding of the structure, organization and evolution of magnetic fields in the very quiet Sun (the regions outside active regions and the network) has become increasingly clear. Magnetic fields in the quietest areas of the Sun are relatively weak and organized at small spatial scales, which yields weak polarization signals that are difficult to observe. Until very recently, the general picture of the structure of its magnetism was rather rough: a ``turbulent'' disorganized field \citep[][]{Stenflo82, Solanki93, rafa_04, TrujilloShchukinaEtal04}.
It is now clear that even in very quiet areas, magnetic fields may organize as coherent loops at granular and subgranular scales \citep[$\lesssim$1000~km;][]{MartinezColladosEtal07}, that these small loops are dynamic \citep{MartinezColladosEtal07, CentenoSocasEtal07, MartinezBellot09, GomoryBeckEtal10},
that they pervade the quiet solar surface and may even connect with upper atmospheric layers \citep{MartinezBellot09, MartinezMansoEtal10}.

Yet, this picture is still incomplete. For example, we lack a complete mapping of the full magnetic field vector on extended fields of view, because the linear polarization signals are intrinsically weak (they are second order on the transverse magnetic field component), and high spatial resolution maps on linear polarization are rather patchy \citep{DanilovicBeeckEtal10}, which has led to incomplete (and sometimes, physically problematic) characterizations of the topology of the field \citep{IshikawaTsunetaEtal08, IshikawaTsuneta09, IshikawaTsuneta10}.

Here we look for and trace small magnetic loops on extended regions of the quiet Sun observed with the highest spatial resolution. Loop-like structures are a natural configuration of the magnetic field due to its solenoidal character. While they can be traced as single, individual, coherent entities, they characterize the magnetic field at large, and their statistics and evolution may shed some light on the origin of the very quiet Sun magnetism, in particular, on the operation or not of local dynamo action \citep{Cattaneo99}. On the other hand, the organization of the field at small scales affect the organization of the magnetic field at larger scales and in higher atmospherics layers \citep{SchrijverTitle03, Cranmer09}, and dynamics \citep{CranmervanBallegoijen10}.

We find evidence for the small scale loops appearing rather irregularly, as in bursts and clumps.
Moreover, wide regions of the very quiet Sun show very low magnetic activity and no apparent sign of organized loops at the detection level of the instruments. These extremely quiet ({\em dead calm}) regions are an intrinsic characteristic of the statistical distribution of these events.

\section{IMaX data}

This paper is devoted to the analysis of disk center quiet Sun observations
obtained with the IMaX instrument \citep{MartinezdelToroEtal11} onboard the SUNRISE
balloon borne observatory \citep{sami_10, barthol_10}. IMaX is a Fabry-P\'erot
interferometer with polarimetric capabilities at the Fe\,{\sc i} line at 5250.2
{\AA}. We analyze two different data sets. Both have the same properties except
that they trace different regions of the quiet Sun and were observed at different times, 
being times series of 22 and 31 min duration. They consist of five filtergrams taken at
$\pm$40, $\pm$80 and +227 m{\AA} from the Fe\,{\sc i} 5250.2 {\AA} line center.
The field of view is 46.8$''\times 46.8''$ (35.6$\times$35.6 Mm$^2$), 20 times larger than the one
observed by \cite{MartinezBellot09}. The spatial resolution is
of about 0$''$.15-0$''$.18$''$. The time cadence is 32 s \citep[note that
it is 28 s in][]{MartinezBellot09},
allowing a noise level of $\sigma= 10^{-3}$ and
$7\times 10^{-4}$ I$_\mathrm{c}$ in the circular and linear polarization, respectively
(I$_\mathrm{c}$ being the continuum intensity).

\begin{figure}[!t]
\includegraphics[width=0.23\columnwidth,bb=30 0 100 341]{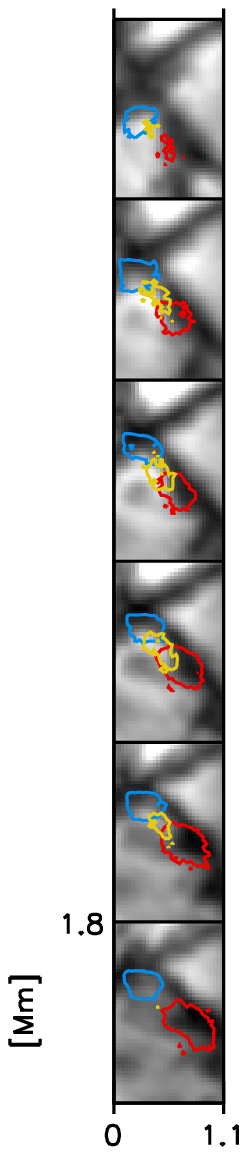}
\hspace{-0.6cm}
\includegraphics[width=0.23\columnwidth,bb=30 0 100 341]{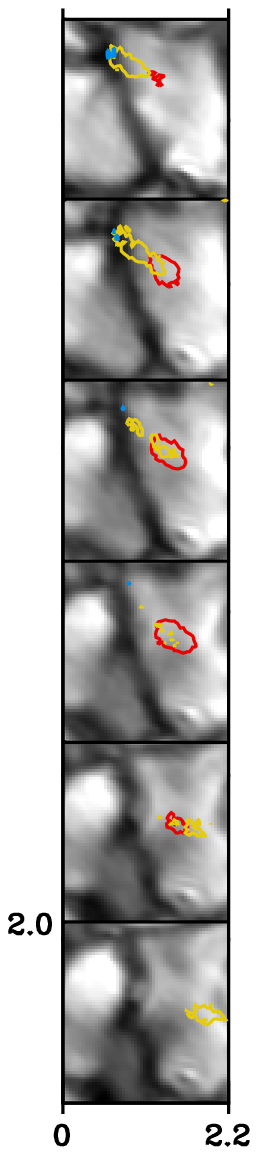}
\hspace{-0.2cm}
\includegraphics[width=0.23\columnwidth,bb=30 0 100 341]{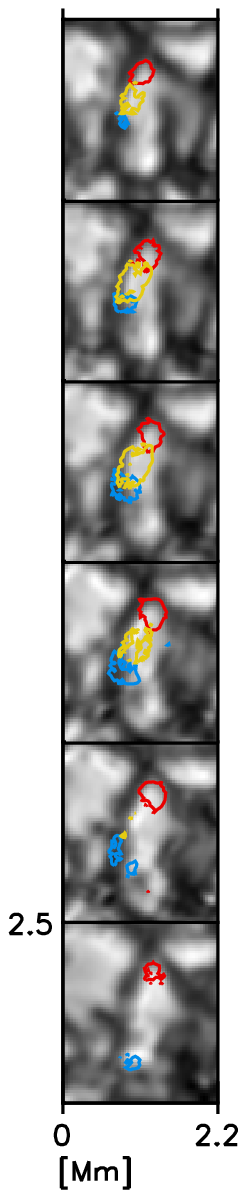}
\hspace{-0.2cm}
\includegraphics[width=0.23\columnwidth,bb=30 0 100 341]{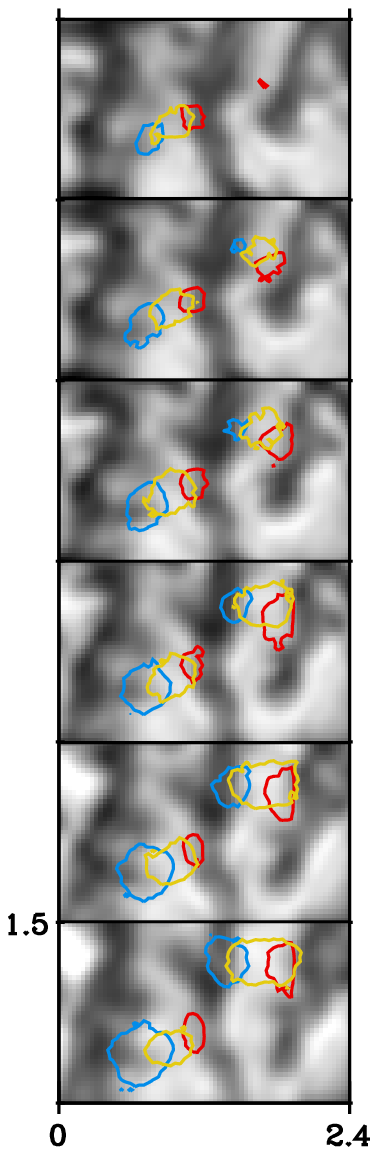}
\caption{Four examples of small loops found in the IMaX data. The grayscale background images represent the 
local granulation. The red and blue contours represent isocontours of magnetic flux density delimiting the positive and negative footpoints of the loop, respectively. The yellow contours represent the linear polarization. The time goes from top to bottom and the time intervals between successive images are not regular. The values of the polarization contours are different but are above the noise level and are chosen in each case to clearly see the loop.}
\label{ejemplos}
\end{figure}

\begin{figure*}[!t]
\includegraphics[width=0.5\textwidth]{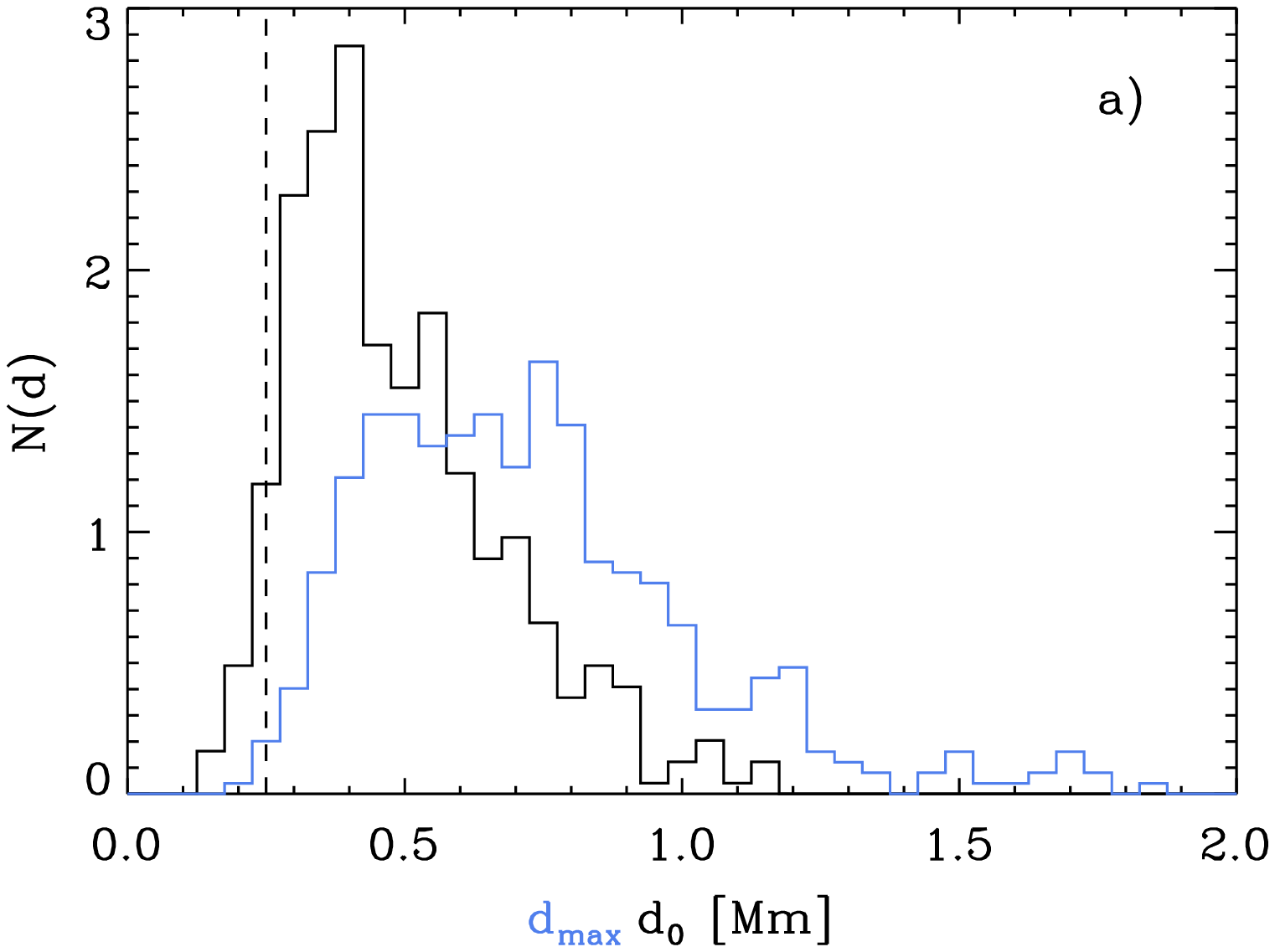}
\includegraphics[width=0.5\textwidth]{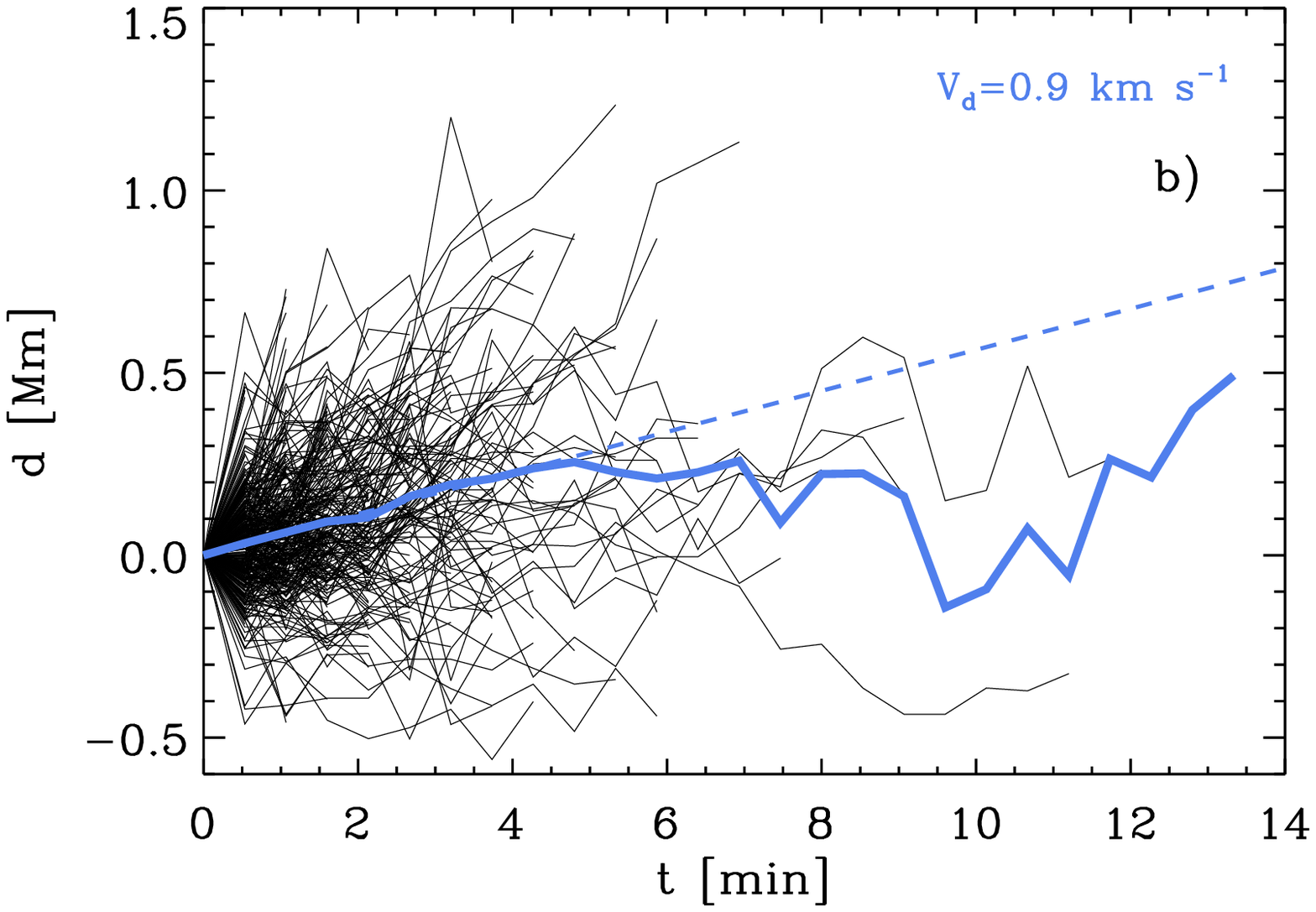}
\includegraphics[width=0.5\textwidth]{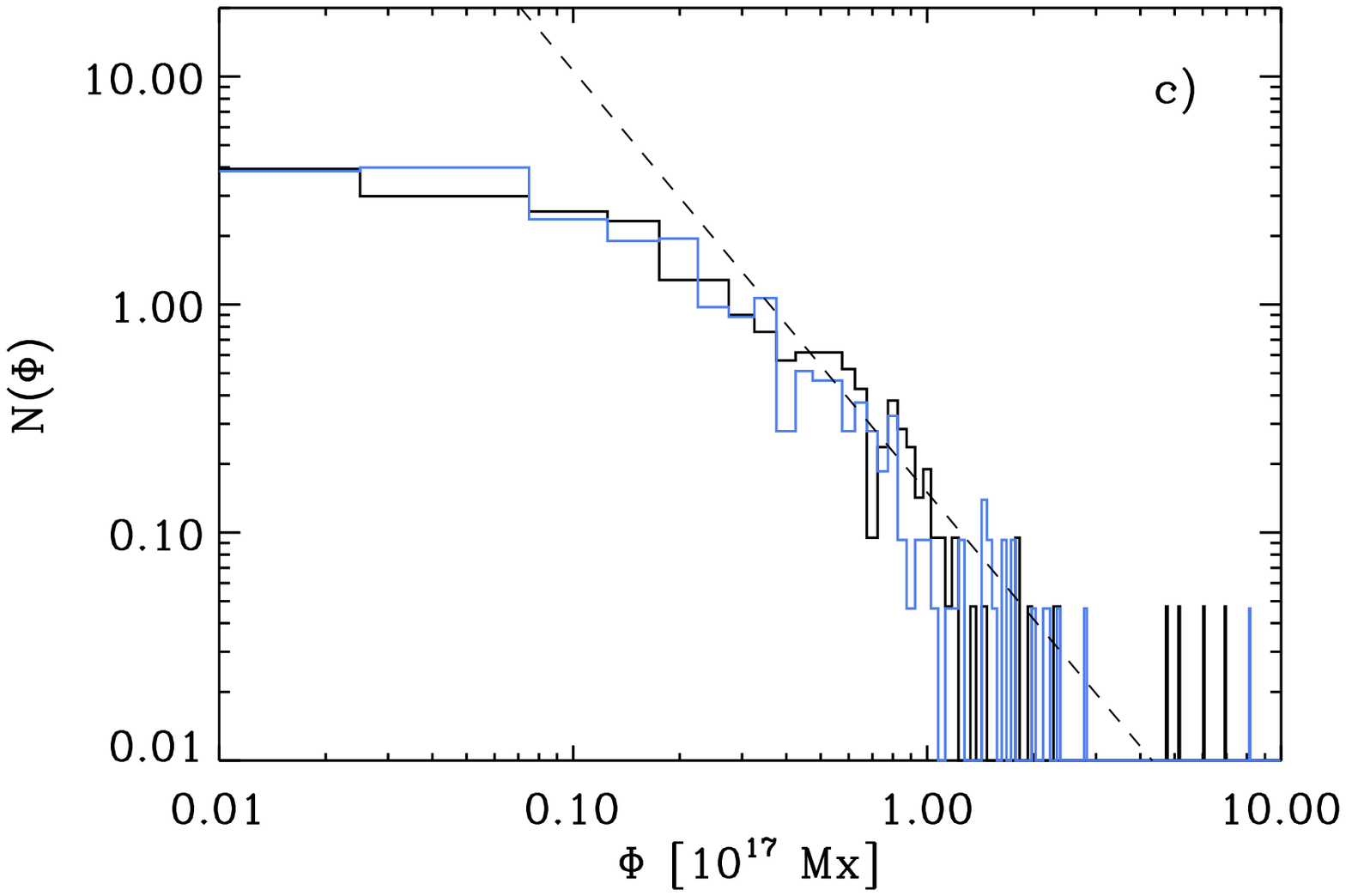}
\includegraphics[width=0.5\textwidth]{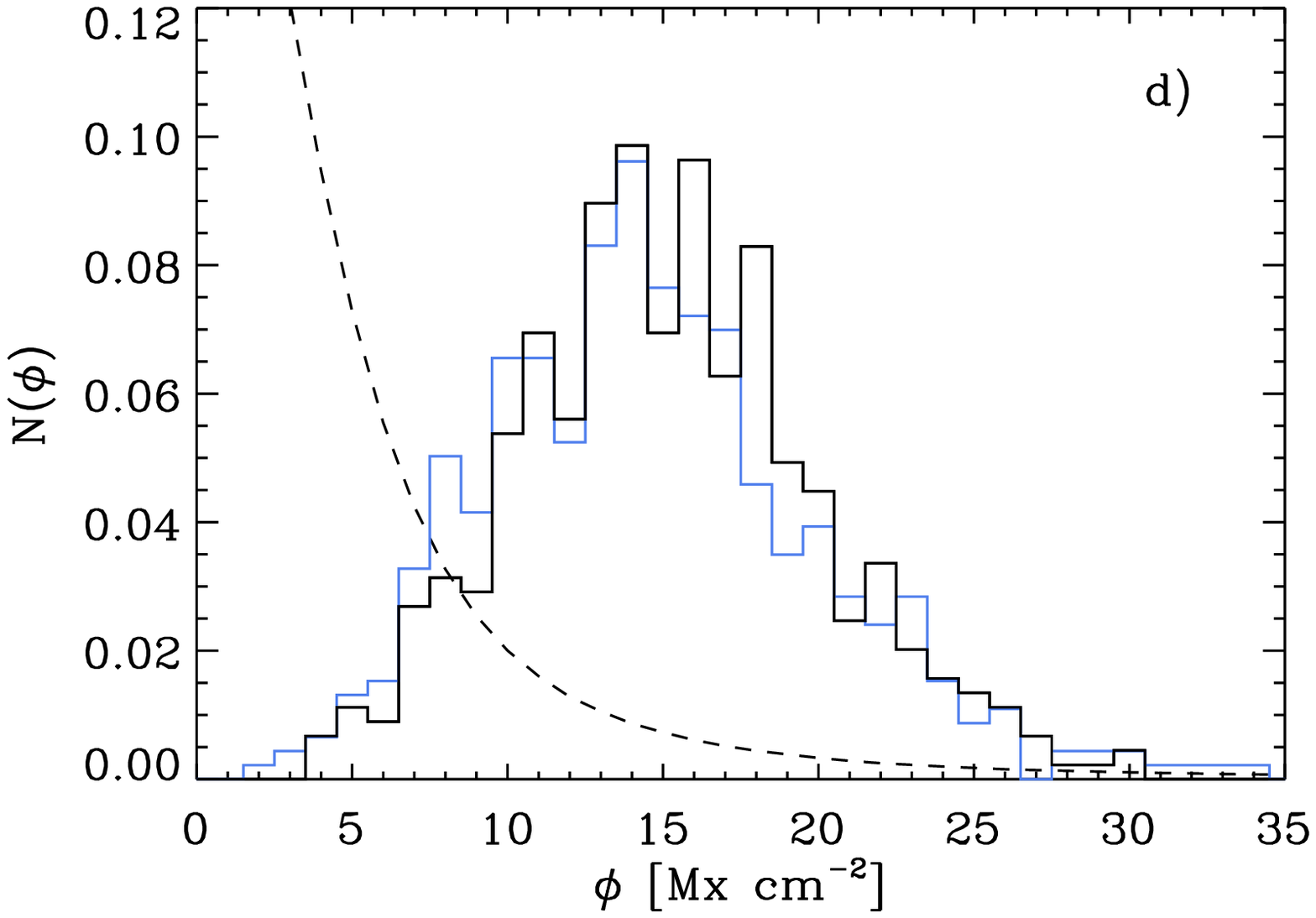}
\caption{Panel a) displays the distribution of the initial distance between
footpoints (black line) and its maximum extent (blue line). The vertical dashed
line represents the mean diameter of the footpoints at the initial time of
detection (when the two footpoints are visible for the first time). Panel b)
displays the time evolution of the footpoints distance with respect to the initial footpoint separation. 
The thick blue line represents the average and the dashed line is the linear fit. From the slope of this fit we 
infer the velocity of footpoint separation of the small dipoles. 
Panel c) and d) represent the histograms of the magnetic flux and the magnetic flux
density of the footpoints of the loops, respectively. The black lines represent the positive values while
the blue lines are the negative ones. The dashed line corresponds to a power law
using the exponent found by \cite{parnell_09}. Note that the proportionality
coefficient is different since the fluxes have been obtained with two different
instruments and have probably an offset.}
\label{magn_loops}
\end{figure*}

\begin{figure*}[!t]
\includegraphics[width=0.5\textwidth, bb= 113 396 419 704]{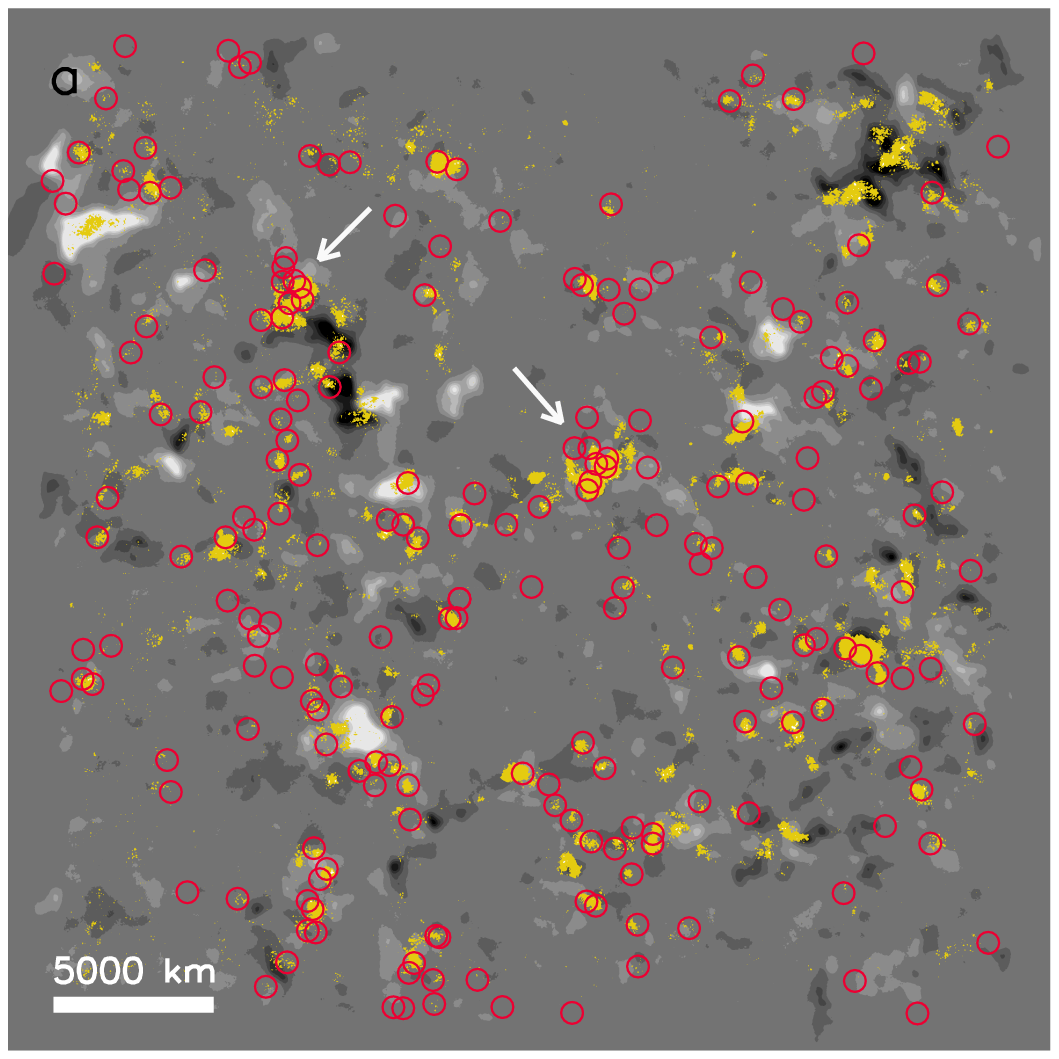}
\includegraphics[width=0.5\textwidth, bb= 113 396 419 704]{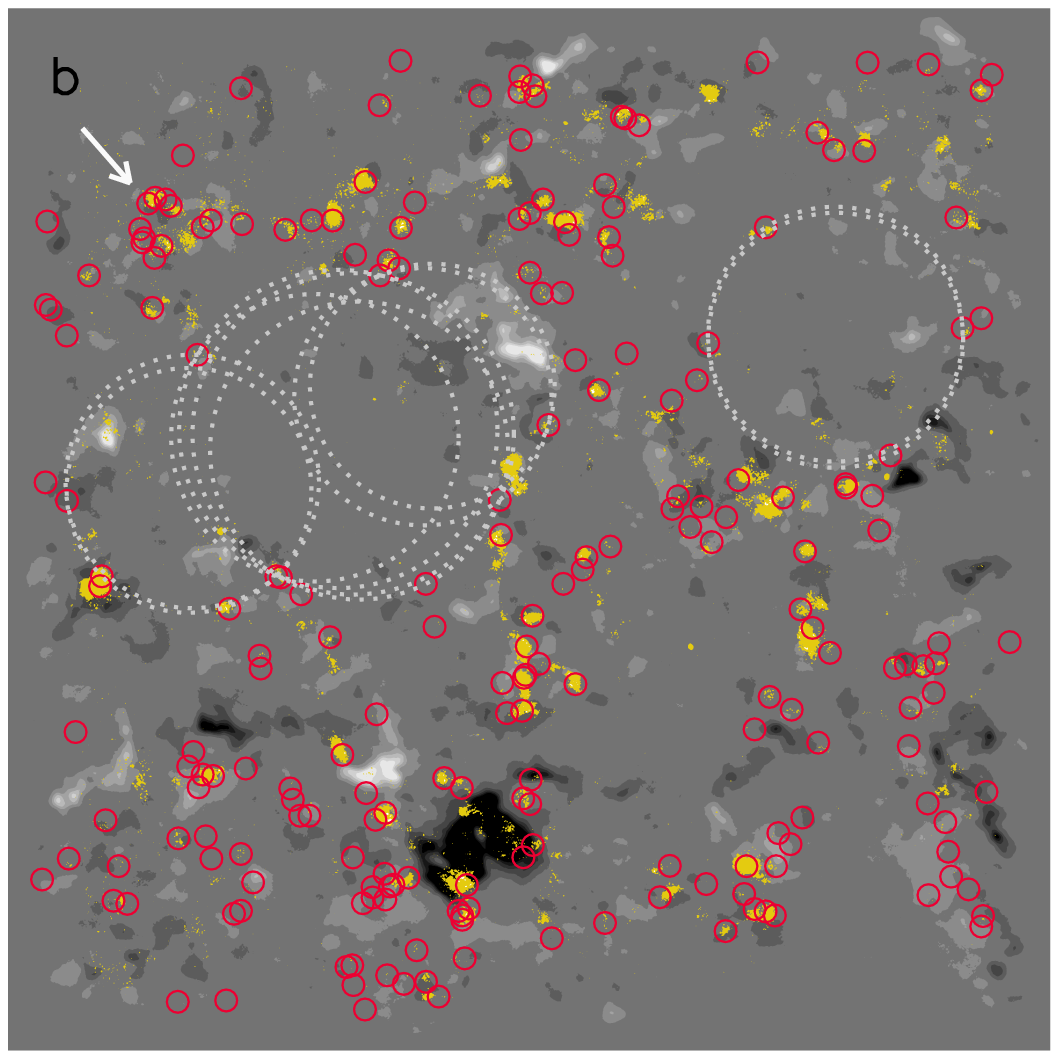}
\caption{Magnetograms of the two areas analyzed in this paper integrated for the whole observational period.
The grey scale has been saturated to show just the most strongly magnetized regions (otherwise, virtually all pixels in the FOV show a detectable signal above noise level).
Yellow patches show the average linear polarization signals after correcting from the bias according to \citep{MartinezMansoEtal12}, and therefore, they are patches with a reliable detection of the transversal component of the magnetic field.
Red circles show the average position (between both footpoints) of the 219 (panel a) and 216 (panel b) small scale loops detected in both datasets. The trajectories they follow while they evolve are comparable to the size of the circles. The two large areas covered by the dotted circles in panel b) mark the two {\em dead calm} regions with a statistically significant lack of (detectable) magnetic loops. Note that they show low magnetic activity too. Arrows indicate a few {\em hotspots} where several loops appear and disappear at approximately the same position, succesively along an extended period of time.}
\label{voids}
\end{figure*}

\section{Small dipoles counting and statistics}

We look for small magnetic loops in the data set:
coherent structures that appear as a dipole in the longitudinal magnetogram (i.e., adjacent positive and negative patches in Stokes-$V$ maps), and a linear polarization patch between them (see Figure 1), and that remain identifiable for several (at least 2) time frames ($>$1 minute).

We looked for small magnetic loops by direct visual inspection. A systematic search was performed in both field-of-views (FOVs) at all times, for these structures by one of the authors (MJMG), recording the position and evolution at different times of every single event. The method was validated by selecting a small area ($10\times 10$~Mm$^2$) and independently looking for such structures by a different observer (RMS). In this control region, the first observer found $N_M=37$ loops, the second one $N_R=40$. $N_{M\cap R}=29$ of them were found by both observers. From those values, the actual number of loops $N$ on the area may be estimated by the Laplace ratio (or Lincoln-Petersen index) for population estimates: $N=N_M N_R/N_{M\cap R}=51$ \citep[e.g.,][]{Cochran78}; a better, less biased, estimate being $N=(N_M+1)(N_R+1)/(N_{M\cap R}+1)-1=50$ \citep{Chapman51}. The variance of 
this estimate is ${\rm Var}(N)\approx(N_M+1)(N_R+1)(N_M-N_{M\cap R})(N_R-N_{M\cap R})/[(N_{M\cap R}+1)^2(N_{M\cap R}+2)]=4.9$ \citep{Chapman51}.

These formulae have been used in the literature to estimate the size of ecological populations \citep[e.g.,][]{Seber02, Krebs08}, and the number of errors in a message \citep[][]{Barrow98}.
The main difficulty for applying the method to our case is that the objects to be counted 
(the loops as defined in the first paragraph of this section) may not be unambiguous.
To guarantee that this condition is fulfilled (i.e., that both observers were identifying and counting the same objects), both observers went again over all the structures found in the area by both, and they had to agree over all the events to be counted as loops. 

From this analysis we conclude that the total number of magnetic structures that we find in the whole 
dataset could be understimated by $\sim$35\%. We note that the most clear cases (those with the strongest polarization signals, lasting longer, and clearly isolated from neighbouring magnetic patches) were often found by both observers in the control area. The 35\% discrepancy is due to those events found by one observer but not the other; these cases correspond to the most subtle (often bordering the detection limit) events. 

We identify 497 small magnetic loops emerging in the observed regions. Taking into account the total time of observation and the spatial area covered, this amounts to an emergence rate of small dipoles of 0.25~loops~h$^{-1}$~arcsec$^{-2}$, rising up to 0.34~loops~h$^{-1}$~arcsec$^{-2}$ when we apply the correction for undetected (although present in the data) loops given above. These values are one order of magnitude larger than the previous estimation found by \cite{MartinezBellot09}, but compatible with Mart\' inez Gonz\'alez (2007) and Mart\' inez Gonz\'alez et al. (2010). There are several reasons for the discrepancy. First, the former study covered a relatively small area and, if the emergence is not strictly uniform (as we shall discuss below), the rate can be greatly understimated. This fact was already pointed out in \cite{MartinezBellot09}, where it was noted that there was evidence for preferential emergence regions. The reason why the studies in the near-IR and this letter are in agreement is because, either having a better Zeeman sensitivity (Mart\' inez Gonz\'alez et al. (2007) and Mart\' inez Gonz\'alez et al. (2010) used the more sensitive-- most importantly to linear polarization-- spectral line of Fe~{\sc i} at 1.5~$\mu$m) or having a better spatial resolution makes both studies more sensitive, inducing the identification of more (weaker and/or smaller) structures in the field.

Both opposite polarity feet and the linearly polarized bridge between them were tracked during the full loop phase. In most cases (60\% of the events), the loop collides or merges to some degree with a neighbouring structure and we could not trace further its individual evolution. This is a common case because at our level of sensitivity and at such spatial resolution, most of the pixels in the FOV show circular polarization, and many, both linear and circular. For the remaining 40\% of the cases, the loop individual history could be traced beyond (and before) their complete loop phase (i.e., when both three polarization patches are seen simultaneously). 
In nearly half of these cases (56\%), both footpoints and linear polarization appear and/or disappear simultaneously because the structure falls below the detection level of the instrument (or it submerges). We named this population of loops as "low-lying" \citep[see][]{MartinezBellot09}. This population of loops is very particular of the quiet Sun. In 42\% of the cases, linear polarization precedes the detection of both footpoints and then disappears before them too. \cite{MartinezBellot09} computed the line-of-sight velocity of the loops using the Stokes $V$ zero-crossing shift and verified that all the loops having this very same time evolution were rising $\Omega$-loops. We have no reasons for thinking that the loops found in IMaX are a different population. However, we cannot compute the Stokes $V$ velocity in the IMaX data sets analyzed in this paper, hence, the identification of these loops as rising $\Omega$-loops can be put in doubt. In just 3 instances (2\%), we found that first the opposite polarities appear, then the linear polarization between them, and everthing disappears, wich could be interpreted as the emergence of a U-loop, or, more probably (considering the local evolution of the flow), a submergence of an $\Omega$-loop.

Figure \ref{ejemplos} display four examples of small loops found in the IMaX data. Note that, for the sake of clarity, we have only drawn the contours of interest, avoiding circular and linear polarization patches adjacent to the loop structure. The time runs from top to bottom, the time cadence being irregular. The first and third columns represent typical loops in which the linear polarization disappears at some point in the evolution of the loop while the footpoints stay in the photosphere (probably rising $\Omega$-loops). These two loops have, however, some peculiarities that differentiate them. The loop 
in the first column appears at the border of an expanding granule. As the granule expands, the entire loop is dragged by the plasma flow. The loop in the third column contains a linear polarization signal with a gap in between. These two peculiarities are a consequence of the spatial resolution of the IMaX data that allows us to trace the dynamics of the linear polarization. 

The loop in the second column of Fig. \ref{ejemplos} has another linear polarization patch with a curious dynamics. It seems that, somehow, the negative footpoint is disconected from the positive one (i.e., the loop breaks) and that it connects somewhere else to the right of the positive footpoint. Of course, it is just a visual impression. The example in the fourth column is an example of two loops appearing very close in time and in space. They also disappear more or less at the same time, hence, one could think this is an evidence of a sea-serpent magnetic field line. Note also that the rightmost loop rotates.

All the observed dipoles are smaller than $\sim 1$~Mm (center-to-center distance between the two opposite polarity patches), becoming increasingly more abundant at smaller scales, with most of the observed dipolar structures being $\sim 0.4$~Mm (Figure 2a). This is barely three times the spatial resolution limit of IMaX, which suggests that the detection and our statistics might be limited by the instrument. The tilt angle of these dipoles is uniformly distributed, meaning that it does not follow the Hale's polarity law. This last result is consistent with \cite{MartinezBellot09} and even with the behaviour of the smallest ephemeral regions. Figure 2b displays the distance between footpoints with respect to the separation between them at the initial time. Therefore, positive values mean that the footpoints separate with time and negative ones indicate that the footpoints approach each other. In average, the distance grows linearly with time with a velocity of $V_d=0.9$ km s$^{-1}$, comparable to typical granular values. This indicates that the loops passively follow the granular flows, as expected from weak magnetic features \citep{rafa_11}. 

The magnetic properties of these small dipoles are represented in Fig. 2c and 2d. They are obtaining inverting the data in the weak field approximation following \cite{MartinezMansoEtal12}. 
The magnetic flux has been computed in the area containing the observed signal. 
The frontier has been defined by eye (as an isocontour of magnetic flux density) and hence is slightly
different for the different structures. The magnetic flux density is the mean
value of the magnetic flux densities in this same region. The magnetic flux of the 
loops can be explained with a power law using the
exponent found by \cite{parnell_09}. This means that the population of small
dipoles follows the population of magnetic fields in the quiet Sun. But looking at 
the histogram of magnetic flux densities, the loops are located at the end tail of the histogram; it is mostly in the range 10-20~Mx~cm$^{-1}$ ---i.e., 10-20~G if the magnetic field were uniformly distributed and volume filling. This is compatible with the results of \cite{marian_rafa_10} who state that, in the quiet Sun, the larger the signal the larger the degree of organization of magnetic fields.

\section{Spatial Distribution}

Although small scale loops are found all over the observed areas, their spatial distribution does not seem to be completely uniform (Fig. 3). It can be observed that, at some locations, loops appear repeatedly and succesively as in bursts, forming clusters, a behaviour that has been noticed before by Mart\' inez Gonz\'alez \& Bellot Rubio (2009), who pointed out that, often, the appearence of loops made it more likely that new ones were later detected nearby.
On the other hand, extended areas seem to be noticeably empty of such events, as if voids appeared in the distribution.
However, this may be deceiving since voids are also formed even in strictly uniform distributions of points \citep{Betancort90, Betancort91}.

A quantitative analysis to determine if these voids are statistically significant was performed.
This requires on the first place, an unambiguous definition of `void', a non-trivial task in itself \citep[see e.g.,][and discussions therein]{KauffmannFairall91, TikhonovKarachentsev06}.
We adopted the simplest definition here and considered only voids of circular shape: the largest empty circle that can be fitted in a given region of the point field ---equivalently, an empty circle limited by three points of the distribution\footnote{This is certainly an overabundant definition: several overlapping circles may be found covering what we intuitively consider as a single ``void''. Algorithms might be devised to merge circles and to find a definition closer the intutitive meaning \citep{Gaite05, Colberg08}. This ``overcounting'' is, however, of no importance for our calculating the probability of finding a void larger than a given size (roughly, all voids are overcounted equally), and we will use this much more simple approach which avoids numerical technicalities.}.

If the loops appeared uniformly on the solar surface, then the probability of finding a void of area between $A$ and $A+dA$ within the field of view (square surface of area $L^2$) would be (see Appendix):
\begin{equation}
P(A)dA=\frac{2}{N_{\rm voids}} (L-r)^2 \frac{(nA)^2}{A} {\rm e}^{-nA}n dA,
\end{equation}
where $A=\pi r^2$, $n$ is the surface density of points, and $N_{\rm voids}$ is the total number of voids espected in the FOV, which is given by Equation~(A1). 
For the two data sets studied here $L=31$~Mm (which is slightly smaller than the nominal FOV because we have excluded the apodized exterior area), and $n=0.22$~Mm$^{-2}$. Note that we only need a single value of $n$ since the number of loops detected in both data sets are very similar (i.e., 248 and 249 events). 
\begin{figure}[!t]
\centerline{\epsfig{figure=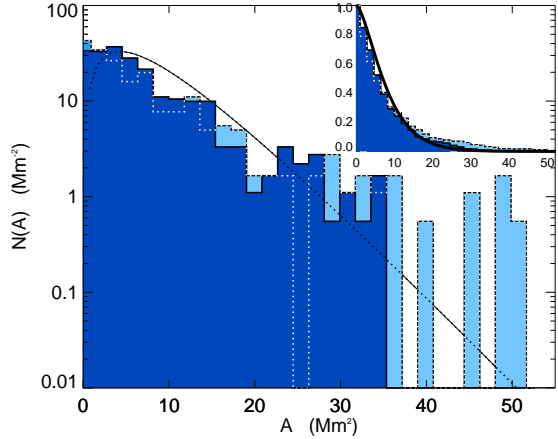, width=8cm}}
\caption{Number of voids per unit area in the two datasets.
Dark blue and cyan histograms correspond to the data shown in Figures 3a and 3b, respectively. 
Solid line shows $N(A)$ for a Poisson distribution of points with $n=0.22$ Mm$^{-2}$ (we use a single value since we observe almost the same number of loops in both data sets, i. e., 248 and 249 loops). The inset window represents the probability of finding an area $\ge A$. The x-axis is the area $A$ in Mm$^2$.}
\end{figure}

Figure 4 shows the number of voids per unit area in both datasets and for the corresponding Poisson distribution.
Except for the smallest areas, the distribution of large circular voids in the first dataset is not significantly different from the uniform one. 
On the second data set, however, apart from the overabundance of small scale voids, large circular voids seem to be significatively more abundant than a strictly uniform distribution would suggest. Actually, for the parameters found for these observations, the probability of finding a circle with an area lager than $A=35$~Mm$^2$ (equivalently, diameter larger than $3.3$~Mm) is very low ($3\times 10^{-4}$; see the inset plot in Fig. 4). We conclude, therefore, that there is statistical evidence for the two voids marked in Figure~3a to be real and not due to chance. Moreover, the oberabundance of small voids is interpreted in terms of a clumpy structuring (see how loops appear in clumps, like a gurgle phenomena, in Fig 3).

In order to relate the distribution of loops (and the voids) to the global magnetism in the observed area, Fig. 3 shows the integrated longitudinal (in black and white) and transverse (in yellow) magnetograms for the two observations. When dealing with the linear polarization, one has to remind that it is a biased estimator of the transversal field component. In the plot, this bias is statistically partially removed as follows: we compute the bias for a percentile 95 when the observations are pure noise \citep[see]{MartinezMansoEtal12} (note that this bias value depend on each pixel, i.e., on the actual intensity profile). This means that the ``real'' transverse magnetic will be below this bias value with a probability of 95\%. We have decided to put all the values smaller than this bias to 0. Figure 3 shows only the statistically significant patches of linear polarization appearing at all the observed times. The positions of the loops do not seem to be clearly correlated with the longitudinal magnetogram, but the voids encircled by the loops show less magnetic activity than other areas in the FOV. Considering the correlation with linear polarization, it suggests that most of the linear polarization signals that are detected are associated to loop structures embeded in the formation region.

\section{Discussion}

It is known that even in very quiet areas of the Sun magnetic fields may organize naturally forming loops at granular scales. In this study we extended this observation to the smallest spatial scales observable (100-1000~km), finding an increasing number of loops at smaller scales up to the resolution limit. 
This finding suggests that the organization of magnetic fields might continue beyond that limit. 
We cannot reconstruct the complete magnetic field topology because 1) the finite spatial resolution of our observations is (perhaps inherently) above the organization scale of the magnetic fields, 
and 2) we lack linear polarimetric sensitivity, which gives us only fragmentary information on the transversal (horizontal) component of the magnetic fields.
Due to these limitations the loop structures that we observe are biased towards relatively large and relatively strong with respect to the magnetic flux density in the neighbouring areas. 
We found evidence that the loops thus detected are not randomly distributed on the solar surface, but rather that they may appear in bursts, and that they are noticiable absent from extended areas which are, also, only weakly magnetized.

It is not yet clear what is the nature of the magnetic fields in the quiet Sun ---what are the fundamental physical mechanisms involved in their generation and evolution. The presence of these {\em dead calm} areas in the quiet Sun (and small scale loops hotspots) represent an important constrain on the origin of magnetic fields in the very quiet Sun and on the dominant dynamic and magnetic mechanisms taking place there. 

It is thought that the magnetism of the quiet Sun can be the result of the emergence of underlying organized magnetic fields \citep{fernando_12} or the dragging of the overlying canopy fields \citep{pietarila+11}. It would then be necessary to understand why there are emergence hotspots and dead calm areas. Another possibility is that they are just recycling of the decay of active regions as they diffuse and migrate to the poles. But it seems unlikely that such random walking would lead to the kind of organized structures and to the spatial patterns reported here. It is also possible that they are linked to some type of dynamo action taking place in the solar surface \citep{Cattaneo99}. It seems now clear that coherent velocity patterns are a requisite for dynamo action to take place \citep{tobias_cattaneo08}. The most obvious coherent velocity pattern in the solar surface is granulation. 
If this velocity pattern is involved in some dynamo action, it is reasonable 
that it forms coherent magnetic structures (such as the loops), although these does not need to be organized at 
the same granular scales; it could well be that they form intermittent patterns as the ones observed here. 
Actually, theoretical considerations \citep{chertkov+99} and laboratory experiments \citep{revelet+08} support the idea that the onset of turbulent dynamo action may be highly intermittent and bursty. Finally, it could just be that these small scale loops represent the far tail of a continous range of structures 
from a global dynamo, just lying at the other end from sunspots. Their spatial statistics would then 
reflect the velocity patterns on the last (shallowest) layers of magnetic field emergence.

Future models that we construct to understand the generation of magnetic fields in the very quiet Sun have to explain the spatio-temporal coherences that we report. Further work is needed to extend these results in larger areas of the Sun and along the solar cycle.

\begin{acknowledgements}
We specially thank Dr. Juan Betancort Rijo for very helpful discussions on the determination of the statistics of voids, which have improved the paper and strengthen the conclusions.  We also thank Valent\' in Mart\' inez Pillet for helpful discussions. Eliot Hijano enjoyed a summer research grant at the Instituto de Astrof\' isica de Canarias. Financial support by the Spanish Ministry of Economy and Competitiveness through projects AYA2010-18029 (Solar Magnetism and Astrophysical Spectropolarimetry) and Consolider-Ingenio 2010 CSD2009-00038 is gratefully acknowledged. 
\end{acknowledgements}

\appendix

\section{Derivation of Equation (1)}

We derive Equation (1) adapting to our case some arguments of the strategy by \cite{Gaite09}.
The probability that three points $i=1, 2, 3$ distributed at random (uniformly) on an square with area $A_t=L^2$, have coordinates between $x_i$ and $x_i+dx_i$, $y_i$ and $y_i+dy_i$ ($0\le x_i\le L$ and $0\le y_i\le L$), is $(1/A_t^3)d^3x_i d^3y_i$. 
With the change of variables $x_i=x_c+r\cos\theta_i$, $y_i=y_c+r\sin\theta_i$, we may express 
the probability for the center of circumcircle of the three points to lie between $x_c$ and $x_c+dx_c$, $y_c$ and $y_c+dy_c$, its radius between $r$ and $r+dr$, and the azimutal angles of each point between $\theta_i$ and $\theta_i+d\theta_i$, then, as $dx_cdy_cdr d^3\theta_i |J(r, \theta_i)|/A_t^3$, where $J(r, \theta_i)=r^3 (\cos\theta_1(\sin\theta_2-\sin\theta_3)+\cos\theta_2(\sin\theta_3-\sin\theta_1)+\cos\theta_3(\sin\theta_1-\sin\theta_2))$ 
 is the determinant of the Jacobian matrix.

The probability of one such a circle to lie between the bounds of the large square and its radius to be between $r$ and $r+dr$ is $P_{\rm in}(r)dr=24\pi(L-r)^2 r^3 dr/A_t^3$, where $x_c$ and $y_c$ have been integrated betwen $r$ and $L-r$, and $\theta_i$ between 0 and 2$\pi$, taking into account that the integral of the angular part of $|J(r, \theta_i)|$ is $24\pi^2$.
Alternatively, the probability density of the area $A=\pi r^2$ of such a circle is $P_{\rm in}(A)dA=12(L-r)^2AdA/A_t^3$.

On the other hand, in a homogeneous Poisson field with density $n=N_t/A_t$ ($N_t$ points randomly distributed over an area $A_t$), the probability of finding $k$ points in a region of area $A$ is $P_k(A)=(nA)^k/k!{\rm e}^{-nA}$.
Therefore, the probability of the circumcircle of three points to be void is $12(L-r)^2 A P_o(A)dA/A_t^3$.

Finally, $N_t$ points determine $N_t(N_t-1)(N_t-2)/3!\approx N_t^3/3!$ ($N_t\gg 1$) different triplets (hence, possible circles), and the number of void circles with areas between $A$ and $A+dA$ within the square bounds is $2(L-r)^2A {\rm e}^{-nA}n^3 dA$.

The total number of void circles is:
\begin{equation}
N_{\rm voids}=\int_0^{N_m} 2(L-r)^2 N{\rm e}^{-N}ndN=\frac{2}{\pi}[2(1+2N_m)-{\rm e}^{-N_m}(2+N_m^2)-3\sqrt{N_m\pi}{\rm erf}[\sqrt{N_m}]],
\end{equation}
where $N_m=n\pi L^2/4$ and $\mathrm{erf}(x)$ is the error function \citep{abramowitz72}.
The probability of voids of area $A$ with our constrains is then the number of voids of area $A$ divided by the total number:
\begin{equation}
P(A)dA=\frac{2}{N_{\rm voids}} (L-r)^2 \frac{(nA)^2}{A} {\rm e}^{-nA}n dA.
\end{equation}

For a large total area $A_t$ ($L, A_{t}, N_m\rightarrow\infty$ at constant $n$), then
$P(A)dA=\frac{(nA)^2}{A}{\rm e}^{-nA}dA$, which coincides with the analysis of \cite{PolitzerPreskill86}.

\begin{acknowledgements}
We specially thank Dr. Juan Betancort Rijo for very helpful discussions on the determination o
f the statistics of voids, which have improved the paper and strengthen the conclusions. The c
ontribution of Eliot Hijano to this work was in the frame of the summer grants offered by the 
Instituto de Astrof\' isica de Canarias. Financial support by the Spanish Ministry of Economy 
and Competitiveness through projects AYA2010-18029 (Solar Magnetism and Astrophysical Spectrop
olarimetry) and Consolider-Ingenio 2010 CSD2009-00038 is gratefully acknowledged. 
\end{acknowledgements}


\end{document}